
\magnification=1200 \vsize=23.5truecm \hsize=16truecm \baselineskip=0.7truecm
\parindent=1truecm \nopagenumbers \font\scap=cmcsc10 \hfuzz=1.0truecm

\null \bigskip  \centerline{\bf DISCRETE PAINLEV\'E EQUATIONS:}
\bigskip  \centerline{\bf COALESCENCES, LIMITS AND DEGENERACIES}
\vskip 2truecm
\centerline{\scap A. Ramani}
\centerline{\sl CPT, Ecole Polytechnique}
\centerline{\sl CNRS, UPR 14}
\centerline{\sl 91128 Palaiseau, France}
\bigskip
\centerline{\scap B. Grammaticos}
\centerline{\sl LPN, Universit\'e Paris VII}
\centerline{\sl Tour 24-14, 5${}^{\grave eme}$ \'etage}
\centerline{\sl 75251 Paris, France}

\vskip 3truecm \noindent Abstract \smallskip
\noindent Starting from the standard form of the five discrete Painlev\'e
equations we show how
one can obtain (through appropriate limits) a host of new equations which
are also the discrete
analogues of the continuous Painlev\'e equations. A particularly
interesting technique is the one
based on the assumption that some simplification takes place in the
autonomous form of the mapping
following which the deautonomization leads to a new $n$-dependence and
introduces more new
discrete Painlev\'e equations.
\vfill\eject

\footline={\hfill\folio} \pageno=2

\noindent 1. {\scap Introduction. }
\smallskip
\noindent The recent intense activity in the domain of integrable discrete
systems [1] has
led to the discovery of these most interesting entities that are the
Painlev\'e mappings
[2]. They are similar in essence to their continuous counterparts. In fact,
for each property
of the continuous Painlev\'e equations there exists a discrete analog [3].
However the
discrete Painlev\'e equations (d-P's) are richer. This becomes manifest
when one examines all
their possible forms. In the continuous case, there exist just one
canonical form for each
Painlev\'e equation, written as a second order differential equation of the
type
$w''=f(w',w,t)$ with $f$ rational in $w'$, algebraic in $w$ and analytic in
$t$. In the discrete
case, on the other hand, there exists a profusion of d-P's.  This is true
even when we make the
restriction to three-point rational mappings, resulting from the
de-autonomization of a Quispel
[4] form
$$x_{n+1}={f_1(x_n)-x_{n-1}f_2(x_n)\over f_4(x_n)-x_{n-1}f_3(x_n)} \eqno(1.1)$$
(where $f_4=f_2$ at the autonomous limit).
No canonical form for the d-P's are known, neither does one know how to
classify them. For historical reasons the basic forms of the first five
d-P's (the form of
d-P$_{\rm VI}$ is unknown to date) are [5]:
$$x_{n+1}+x_{n-1}=-x_n+{z\over x_n}+a\eqno (1.2a)$$
$$x_{n+1}+x_{n-1}={zx_n+a\over 1-x_n^2}\eqno (1.2b)$$
$$x_{n+1}x_{n-1}={ab(x_n-p)(x_n-q)\over (x_n-a)(x_n-b)}\eqno (1.2c)$$
$$(x_{n+1}+x_n)(x_n+x_{n-1})={(x_n^2-a^2)(x_n^2-b^2)\over
(x_n-z)^2-c^2}\eqno (1.2d)$$
$$(x_{n+1}x_n-1)(x_nx_{n-1}-1)={pq(x_n-a)(x_n-1/a)(x_n-b)(x_n-1/b)\over(x_n-
p)(x_n-q)}\eqno(1.2e)$$
where $z=\alpha n+\beta$, $p=p_0\lambda^n$, $q=q_0\lambda^n$ and $a$, $b$,
$c$ constants.

In the continuous case, the six Painlev\'e equations are known to form a
coalescence
cascade [6]. This means that by taking the appropriate limits of the
dependent and independent
variables $(w,t)$ as well as the parameters of the equation, we can recover
a `lower' equation
starting form a `higher' in the following pattern:

P$_{\rm VI}$ $\to$ P$_{\rm V}$ $\to$ $\{$P$_{\rm IV}$, P$_{\rm III}\}$
$\to$ P$_{\rm
II}$ $\to$ P$_{\rm I}$

\noindent We can easily show that this is true for the discrete equations
as well (see Section 2).
However, the d-P's are related by more than just the coalescence cascade.
In order to  fix the
ideas let us summarize here what we mean by coalescences, limits and
degeneracies. Coalescence is
a limiting procedure performed on the parameters of the equation but also
on the dependent
variable as well as on the explicitly $n$-dependent ones. In that way one
gets an equation that
has fewer parameters than the equation one starts with. The limits
correspond to taking some
parameters of the equation to zero or infinity. The remaining parameters
have the {\sl same}
$n$-dependence as in the initial equation. In this way one obtains either a
special case of the
initial equation (a trivial case) or a new equation with fewer parameters.
Finally, the
degenerate forms are obtained if one assumes that some simplification
occurs to the initial
equation {\sl prior to deautonomization}. Due to this simplification, we
may find {\sl new}
deautonomizations (corresponding to equations with fewer parameters than
the initial one)
leading to an equation the form of which cannot be retrieved from the
original neither as a limit
nor as a coalescence. This procedure allows us to generate new discrete
Painlev\'e equations.

In what follows, we shall study the limits and degeneracies of the five
`standard' d-P's (1.2).
Whenever an interesting  degenerate form is obtained in the autonomous
case, the deautonomization
is performed on the basis of the singularity confinement criterion [7].
Namely, we accept only
those nonautonomous forms that satisfy this discrete integrability detector
that we have
developed in the recent past and which has proven already to be an
efficient and valuable tool
[8].
\bigskip
\noindent 2. {\scap Coalescence cascade of the discrete Painlev\'e equations. }
\smallskip
\noindent As we have explained in the introduction, the d-P's form a
coalescence cascade allowing
one to obtain `lower' ones starting from a `higher' one by taking the
appropriate limits of
dependent variables as well as of the parameters (and also the explicitly
$n$-dependent
variables). The analogy with the continuous Painlev\'e equations is
perfect. In this case the
coalescence chain is:

d-P$_{\rm V}$ $\to$ $\{$d-P$_{\rm IV}$, d-P$_{\rm III}\}$ $\to$ d-P$_{\rm
II}$ $\to$ d-P$_{\rm I}$

\noindent In what follows, we will present the result for the
five standard forms (1.2a-e). The following conventions will be used. The
variables and
parameters of the `higher' equation will be given in capital letters
($X,Z,P,Q,A,B,C$), while those of the `lower' equation are given in
lowercase letters
($x,z,p,q,a,b,c$). The small parameter that will introduce the coalescence
limit will be
denoted by $\delta$.

In order to illustrate the process, let us work out in full detail the case
d-P$_{\rm
II}$ $\to$ d-P$_{\rm I}$. We start with the equation:
$$X_{n+1}+X_{n-1}={ZX_n+A\over 1-X_n^2} \eqno(2.1)$$
We put $X=1+\delta x$ whereupon the equation becomes:
$$4+2\delta(x_{n+1}+x_{n-1}+x_n)=-{Z(1+\delta x_n)+A\over \delta x_n}
\eqno(2.2)$$
Now, clearly, $Z$ must cancel $A$ up to order $\delta$ and this suggests
the ansatz
$Z=-A-2\delta^2z$. Moreover, the ${\cal O}(\delta^0)$ term in the rhs must
cancel the 4 of the lhs
and we are thus led to $A=4+2\delta a$. Using these values of $Z$ and $A$
we find (at $\delta\to
0$):
$$x_{n+1}+x_{n-1}+x_n={z\over x_n}+a \eqno(2.3)$$
i.e. precisely d-P$_{\rm I}$.

The coalescence d-P$_{\rm III}$ to d-P$_{\rm II}$ requires a more delicate
limit since the
independent variable of d-P$_{\rm III}$ enters in an exponential way. In
order to perform the
limit we take $\lambda=1+\gamma\delta^r$ for some $r$, whereupon
$\lambda^n$ becomes
$1+n\gamma\delta^r+{\cal O}(\delta^{2r})$  and thus, at the limit, $p,q$
are of the form $\alpha
+\beta n+{\cal O}(\delta^{2r})$ with $\beta=\alpha\gamma\delta^r$ . We start
from:
$$X_{n+1}X_{n-1}={AB(X_n-P)(X_n-Q)\over (X_n-A)(X_n-B)} \eqno(2.4)$$
The ansatz for $X$ is here, too, $X=1+\delta x$. For the remaining
quantities we find:
$$A=1+\delta , \quad B=1-\delta$$
$$P=1+\delta+\delta^2(z+a)/2+{\cal O}(\delta^{3}) ,\quad
Q=1-\delta+\delta^2(z-a)/2+{\cal O}(\delta^{3})\eqno(2.5)$$
so in fact $r=2$, and at the limit $\delta\to 0$, d-P$_{\rm III}$ reduces
exactly to d-P$_{\rm
II}$:
$$x_{n+1}+x_{n-1}={zx_n+a\over 1-x_n^2} \eqno(2.6)$$
As we saw above, d-P$_{\rm IV}$ also reduces to d-P$_{\rm II}$. Here we
start from:
$$(X_{n+1}+X_n)(X_n+X_{n-1})={(X_n^2-A^2)(X_n^2-B^2)\over (X_n-Z)^2-C^2}
\eqno(2.7)$$
and put $X=1+\delta x$. We take:
$$A=1+\delta ,\quad B=1-\delta$$
$$C=\delta-\delta^2a/2 ,\quad Z=1-\delta^2z/4\eqno(2.8)$$
The result at $\delta\to 0$ is precisely d-P$_{\rm II}$ given by (2.6).

In the case of d-P$_{\rm V}$:
$$(X_{n+1}X_n-1)(X_nX_{n-1}-1)={PQ(X_n-A)(X_n-1/A)(X_n-B)(X_n-1/B)\over
(X_n-P)(X_n-Q)}
\eqno(2.9)$$
two different limits exist. In order to obtain d-P$_{\rm IV}$ we put
$X=1+\delta x$ and take:
$$A=1+\delta a ,\quad B=1-\delta b$$
$$P=1+\delta (z+c) ,\quad Q=1+\delta (z-c)\eqno(2.10)$$
i.e. $\lambda=1+\alpha\delta$, such that $z=\alpha n+\beta$.
At the limit $\delta\to 0$ we find d-P$_{\rm IV}$ (1.2d) in terms of the
variable $x$. The case of
the coalescence d-P$_{\rm V}$ to d-P$_{\rm III}$ requires a different
ansatz. Here we put
$X=x/\delta$. Moreover we take:
$$P={p\over \delta},\quad Q={q\over \delta},\quad
A={a\over \delta},\quad  B={b\over \delta}\eqno(2.11)$$
We find then at the limit $\delta\to 0$:
$$x_{n+1}x_{n-1}={pq(x_n-a)(x_n-b)\over (x_n-p)(x_n-q)} \eqno(2.12)$$
While this is not exactly the form of d-P$_{\rm III}$ (1.2c or 2.4) it is
very easy to reduce it
to the latter. We introduce $y$ through $x=y\lambda^n$ (recall
$p=p_0\lambda^n$, $q=q_0\lambda^n$)
and find with $\mu=1/\lambda$:
$$y_{n+1}y_{n-1}={p_0q_0(x_n-a\mu^{n})(x_n-b\mu^n)\over (x_n-p_0)(x_n-q_0)}
\eqno(2.13)$$
that is obviously of the form (2.4).

While, as far as coalescence is concerned, the discrete Painlev\'e
equations follow closely the
behaviour of the continuous ones, this will not be the case of limits and
degeneracies. As we
will see in the following sections, the d-P's have a very rich structure.
\bigskip
\noindent 3. {\scap Limits and degenerate forms of the }d-P$_{\rm
I}$/d-P$_{\rm II}$
{\scap equations. }
\smallskip
\noindent In this section we shall examine the possible forms of discrete
Painlev\'e equations
that have the same $x_{n+1}, x_{n-1}$ dependence as d-P$_{\rm I}$and
d-P$_{\rm II}$, namely:
$$x_{n+1}+x_{n-1}=-{\beta x_n^2+\epsilon x_n+\zeta\over \alpha x_n^2+\beta
x_n+\gamma}\eqno(3.1)$$
 In (3.1) the standard notations of the Quispel mapping have been used. The
singularity
confinement integrability criterion can be used on (3.1) in order to
determine its possible
deautonomizations. Two case must be distinguished from the outset.

\noindent a) d-P$_{\rm I}$, correponding to $\alpha=0$ in which case we can
take $\beta=1$ and
$\gamma=0$ through a simple translation, (the case $\alpha=\beta=0$ being
linear, thus trivial).
The deautonomization of this case leads simply to:
$$x_{n+1}+x_{n-1}+x_n={z\over x_n}+a \eqno(3.2)$$
where $z$ is linear in $n$, $z=\lambda n+\kappa$, and it is just the
`standard' d-P$_{\rm I}$. The
limit $a=0$ of (3.2) was examined in [9] and dubbed d-P$_{\rm 0}$. The
latter is an equation that
does not possess any interesting continuous limit.

\noindent b) d-P$_{\rm II}$, correponding to $\alpha\not = 0$, in which
case we can take
$\alpha=1$ and $\beta=0$ by translation. Two cases can be distinguished,
both with $\epsilon$
linear in $n$. The first, $\gamma=-1$ is just  d-P$_{\rm II}$:
$$x_{n+1}+x_{n-1}={zx_n+a\over 1-x_n^2}\eqno (3.3)$$
 while the limit $\gamma\to 0$ corresponds to a known form [10] of
d-P$_{\rm I}$:
$$x_{n+1}+x_{n-1}={z\over x_n}+{a\over x_n^2}\eqno (3.4)$$
The analysis above has dealt with the limits of  d-P$_{\rm I-II}$. However,
another possibility
exists. Suppose that the numerator and denominator in the rhs of (3.1) have
a common factor. This
is what we call degenerate case. This case is of interest only when
$\alpha\not =0$ (otherwise the
degenerate equation becomes linear). In this case we obtain:
$$x_{n+1}+x_{n-1}={\epsilon\over x_n+\rho}\eqno (3.5)$$
We can translate $\rho$ to zero and deautonomizing (3.5), using singularity
confinement, we
obtain:
$$x_{n+1}+x_{n-1}={z\over x_n}+a\eqno (3.6)$$
 with again $z$ linear in  $n$ and $a$ constant, which is another form of
d-P$_{\rm I}$ [5].
Its continuous limit is obtained through
$x=1+\epsilon^2w$, $a=4$, $z=-2-\epsilon^5n$, leading at $\epsilon\to 0$ to
$w''+2w^2+t=0$, with
$t=\epsilon n$. (The same convention $t=\epsilon n$ will be used throughout
this paper).
\bigskip
\noindent 4. {\scap Limits and degenerate forms of the }d-P$_{\rm III}$
{\scap equation. }
\smallskip
\noindent Let us start with the form of d-P$_{\rm III}$ obtained by
deautonomization of the
Quispel mapping with $f_2=0$ :
$$x_{n+1}x_{n-1}=-{\gamma x_n^2+\zeta x_n+\mu\over \alpha x_n^2+\beta
x_n+\gamma}\eqno(4.1)$$
The full d-P$_{\rm III}$ corresponds to $\gamma\not = 0$. We find (through
application of the
singularity confinement criterion) that $\zeta=\zeta_0\lambda^n$
and$\mu=\mu_0\lambda^{2n}$.
Special values of $\alpha$, $\beta$, $\zeta$ and $\mu$ just lead to special
forms of d-P$_{\rm
III}$ with less than the full complement of free parameters.  For instance,
the limit $\alpha=0$
in (4.1) does not present any particular interest and the transformation
$x\to 1/x$ reduces the equation to d-P$_{\rm III}$ with $\mu=0$. In the
case $\alpha=\beta=0$ we
find (through $x\to 1/x$) a d-P$_{\rm III}$ with $\mu=\zeta=0$ which,
moreover, is strictly
autonomous.
The limits can be readily obtained. When $\gamma=0$ we find the
equation (still for $\zeta\propto\lambda^n,\mu\propto\lambda^{2n}$):
$$x_{n+1}x_{n-1}={\zeta x_n+\mu\over (x_n+\beta) x_n}\eqno(4.2)$$
This is a novel form of d-P$_{\rm II}$. Its continuous limit can be
obtained through
$x=1+\epsilon w$, $\beta=-2+\epsilon^3g$, $\zeta=-2\lambda^n$,
$\mu=\lambda^{2n}$ where
$\lambda=1+\epsilon^3/2$, leading to $w''=2w^3+wt+g$. A complete study of
this equation: special
solutions, B\"acklund and Miura transforms etc. is reserved for a future
publication [11].
A further limit can be obtained, starting from (4.1), by taking
$\beta=0$, in addition to $\gamma=0$. In this case we find:
$$x_{n+1}x_{n-1}={\zeta \over x_n}+{\mu\over x_n^2}\eqno(4.3)$$
where, by the gauge $x\to x\lambda^{n/2}$, $\mu$ can be taken as a constant
and $\zeta$ of the
form  $\zeta_0\lambda^{n/2}$. Equation (4.3) is a discrete P$_{\rm I}$ [12]
as can be seen from
the continuous limit obtained through
$x=1+\epsilon^2 w$, $\zeta=4\kappa^n$, $\mu=-3$ and
$\kappa(\equiv\lambda^{1/2})=1-\epsilon^5/4$,
leading to
$w''=6w^2+t$.

Let us now consider the degenerate forms of (4.1). They are obtained when the
numerator and the denominator in the rhs of (4.1) have a common factor. We
have in
this case:
$$x_{n+1}x_{n-1}={ax_n+b \over c x_n+d}\eqno(4.4)$$
The deautonomization of this equation yields $a=a_0\lambda^n$ and
$d=d_0\lambda^n$. Unless $c=0$,we can always take $c=1$, through division,
and a proper gauge
allows us to take $b=1$. Equation (4.4) in its nonautonomous form is a
novel form
of discrete  P$_{\rm II}$ [11]. The limit $d=0$ in (4.4) leads to the
equation ($c=1$):
$$x_{n+1}x_{n-1}=a+{1 \over x_n}\eqno(4.5)$$
where $a=a_0\lambda^n$. This is another form of d-P$_{\rm I}$. The
continuous limit is obtained
through $x=x_0(1+\epsilon^2 w)$ where $x_0^3=-1/2$, $a=3x_0^2\lambda^n$, with
$\lambda=1-\epsilon^5/3$, leading to
$w''+3w^2+t=0$.
An equivalent equation can be obtained from (4.4) by taking $a=0$:
$$x_{n+1}x_{n-1}={1 \over x_n+d}\eqno(4.6)$$
Equation (4.6) is transformed into (4.5) by taking $x\to 1/x$ and
exchanging $a,d$.
Another limit, leading to another d-P$_{\rm I}$, is $c=0$. We find:
$$x_{n+1}x_{n-1}=ax_n+b\eqno(4.7)$$
where $a$, here, is a constant and $b=b_0\lambda^n$
with continuous limit $w''+6w^2+t=0$ obtained through
$x=1+\epsilon^2 w$, $a=2$, $b_0=-1$ and $\lambda=1+\epsilon^5$. An
equivalent equation can be
obtained also by taking $b=0$ in (4.4). We find:
$$x_{n+1}x_{n-1}={ax_n \over x_n+d}\eqno(4.8)$$
Equations (4.8) and (4.7) are related through the transformation
$x\to 1/x$, with the appropriate relations of the parameters.

We remark that d-P$_{\rm III}$ is particularly rich, since it has yielded
two new
multiplicative d-P$_{\rm II}$'s and five d-P$_{\rm I}$'s, three of which are
genuinely independent forms.
\bigskip
\noindent 5. {\scap Limits and degenerate forms of the }d-P$_{\rm IV}$
{\scap and }
d-P$_{\rm 34}$ {\scap equations.}
\smallskip
\noindent The fact that we treat d-P$_{\rm 34}$ as a fundamental equation
should not be considered a curiosity. Just as in the continuous case
(number 34 in the Gambier
classification), this equation is of capital importance.
It is, in fact, the `modified' d-P$_{\rm II}$, in the sense that it is related
to d-P$_{\rm II}$ through a Miura transformation [13] in perfect analogy to the
continuous case. Given the $x_{n+1}, x_{n-1}$ dependence of d-P$_{\rm 34}$ (we
have in fact $f_2 = -x f_3$ in the Quispel notations (1.1)) it is quite natural
to consider this equation together with d-P$_{\rm IV}$.

Both  d-P$_{\rm IV}$ and d-P$_{\rm 34}$ start from a Quispel form that can be
written as:
$$(x_{n+1}+x_n)(x_n+x_{n-1})= {\alpha x_n^4+\kappa x_n^2+\mu\over \alpha
x_n^2+\beta
x_n+\gamma}\eqno(5.1)$$
a) Let us start with the study of d-P$_{\rm 34}$ which corresponds to
$\alpha=0$ [10].
Its standard form is obtained for $\beta\not=0$ (and we take $\beta=1$):
$$(x_{n+1}+x_n)(x_n+x_{n-1})= {\kappa x_n^2+\mu\over x_n+\gamma}\eqno(5.2)$$
The deautonomization yields $\gamma=z$, and constant $\kappa, \mu$ as only
singularity confining
case. Special limits can be obtained. The simplest is the one for $\kappa=0$:
$$(x_{n+1}+x_n)(x_n+x_{n-1})= {\mu\over x_n+z}\eqno(5.3)$$
This is a form of d-P$_{\rm I}$. However this is not a new d-P$_{\rm I}$ as
can be
seen through the following transformation [10]. We put $y=1/(x_n+x_{n-1})$
and finally obtain
for $y$ the d-P$_{\rm I}$ (3.4). A much more interesting limit is the one
corresponding to
$\alpha=\beta=0$ in (5.1). We have now:
$$(x_{n+1}+x_n)(x_n+x_{n-1})= a_n(x_n^2-b^2)\eqno(5.4)$$
The singularity confinement criterion leads to $b$ constant and $a$ a free
function of $n$! From
our experience on integrable mappings we expect (5.4) to be integrable through
linearization [14]. This turns out to be true. It can be shown that the
solution of
(5.4) is obtained by solving first $(y_{n+1}+1)(y_n-1)=-4/a_n$ where $y$ is
related to $x$
through
$y_n(x_{n-1}+x_n)+x_n-x_{n-1}-2b=0$. Thus this mapping comes from the
coupling of a linear to a
Riccati (homographic) mapping. The continuous limit is consistent with this
result. Putting
$x=1+\epsilon^2 w$, $b^2=\epsilon^3/4$, $a=4+2\epsilon^2 f(\epsilon n)$,
(we recall that we
have a standing convention $t=\epsilon n$),  at the
$\epsilon\to 0$ limit we find that (5.4) is a discretization of (a
particular case of) the
Gambier equation (i.e. equation number 27 in the Painlev\'e-Gambier
classification) [15]:
$w''={w'^2\over 2w}+wf(t)-{1\over 2w}$.

A degenerate case of (5.2) exists when the numerator and the denominator of the
rhs of (5.2) have a common factor. In this case we find:
$$(x_{n+1}+x_n)(x_n+x_{n-1})= ax_n+b\eqno(5.5)$$
The deautonomization of (5.5) leads to a constant $a$ and $b=z$. This
equation is
also a d-P$_{\rm I}$ although not a new one. Putting $y_n=x_{n+1}+x_n$ we find
indeed for $y$ the equation $y_{n+1}+y_{n-1}=a+(z_n+z_{n+1})/y_n$ i.e. equation
(3.6).

\noindent b) we now turn to the full d-P$_{\rm IV}$ i.e. $\alpha\not=0$
that we rewrite as:
$$(x_{n+1}+x_n)(x_n+x_{n-1})= {(x_n^2-a^2)(x_n^2-b^2)\over
(x_n-p)(x_n-q)}\eqno(5.6)$$
The full d-P$_{\rm IV}$ corresponds to $p=z+c,\ q=z-c$. The limiting cases of
d-P$_{\rm IV}$ do not present any interest. Thus, we look directly at the
degenerate
cases. First we have the case of a rhs of (5.1) with cubic numerator and linear
denominator. The application of singularity confinement criterion yields
{\sl two}
different d-P's. We have:
$$(x_{n+1}+x_n)(x_n+x_{n-1})= {(x_n+z)(x_n^2-b^2)\over (x_n-z)}\eqno(5.7)$$
which is yet another novel form of  d-P$_{\rm I}$. Its continuous limit can
be obtained through
$x=5+\epsilon^2 w$, $b^2=-375$ and $z=-3+\epsilon^5 n$ leading to
$4w''+3w^2-25t=0$. But, besides
(5.7), we find also:
$$(x_{n+1}+x_n)(x_n+x_{n-1})= {(x_n+z+k)(x_n^2-b^2)\over (x_n-2z)}\eqno(5.8)$$
which turns out to be a novel form of d-P$_{\rm II}$. Its continuous limit
is given
by $x=6+\epsilon w$,  $b=18$, $k=-9+g\epsilon^3$  and   $z=7+\epsilon^3 n$
leading to
$32w''=w^3-36wt+96g$. Finally
the doubly degenerate case leads, after deautonomozation, to:
$$(x_{n+1}+x_n)(x_n+x_{n-1})= (x_n-z)^2-c^2\eqno(5.9)$$
another form of d-P$_{\rm I}$ with continuous limit $x=1+\epsilon^2 w$,
$c^2=-12$,
and   $z=-3+\epsilon^5 n$ leading to
$w''+3w^2/2+4t=0$.

\bigskip
\noindent 6. {\scap Limits and degenerate forms of the }d-P$_{\rm V}$ {\scap
equation.}
\smallskip
\noindent We shall conclude our study with d-P$_{\rm V}$. Since this is the
equation with the largest number of parameters (among the ones studied here) we
expect its limits to be particularly rich. In order to study these limits it is
more convenient to start with a form:
$$(x_{n+1}x_n-1)(x_nx_{n-1}-1)={\gamma (x_n^4+1)+\kappa x_n(x_n^2+1)+\mu x_n^2
\over \alpha x_n^2+\beta x_n+\gamma}\eqno(6.1)$$
{}From [5,12] we know that the nonautonomous form of d-P$_{\rm V}$ corresponds
to
$\alpha$=constant, $\beta\propto\lambda^n$ and
$\gamma,\kappa,\mu\propto\lambda^{2n}$. The limits of $\alpha$ or $\beta$
equal to zero with $\gamma\not=0$ do not present any interest: they
correspond to
particular cases of d-P$_{\rm V}$. The interesting case is $\gamma=0$. We
have then:
$$(x_{n+1}x_n-1)(x_nx_{n-1}-1)={\kappa (x_n^2+1)+\mu x_n\over \alpha
x_n+\beta}\eqno(6.2)$$
This equation is a novel form of d-P$_{\rm IV}$ as can be seen through the
continuous limit $x=1+\epsilon w$, $\alpha=1+\epsilon^2 a$,
$\beta=-2\lambda^{n}$,
$\kappa=-4\lambda^{2n}(1-\epsilon^4 g)$, $\mu=8\lambda^{2n}$ where
$\lambda=1-\epsilon^2$,
leading to
$w''={w'^2\over 2w}+{3\over 2}w^3+4w^2t+2w(t^2+a)+{g\over w}$.

If, moreover we put another parameter to zero the equation
reduces further to:

\noindent i) if $\alpha=0$
$$(x_{n+1}x_n-1)(x_nx_{n-1}-1)={1\over \beta}(\kappa (x_n^2+1)+\mu x_n)$$
that can be written more conveniently as:
$$(x_{n+1}x_n-1)(x_nx_{n-1}-1)=\kappa(x_n-a)(x_n-1/a)\eqno(6.3)$$
where $\kappa\propto\lambda^n$ and $a$ is a constant.
This is a new form of d-P$_{\rm 34}$ (continuous limit:
$w''={w'^2\over 2w}-2w^2-wt-{2g^2\over w}$
obtained through $x=1+\epsilon^2 w$, $a=1+\epsilon^3 g$, $\kappa=4\lambda^{n}$,
$\lambda=1-\epsilon^3/2$).

\noindent ii) If $\kappa=0$
$$(x_{n+1}x_n-1)(x_nx_{n-1}-1)={\mu x_n\over \alpha x_n+\beta}\eqno(6.4)$$
which is a new d-P$_{\rm II}$ (continuous limit:
$w''+2w^3-2wt+p=0$ obtained through $x=i+\epsilon w$, $\alpha=1+i\epsilon^3
p/2$,
$\beta=-2i\lambda^{n}$,
$\mu=-4\lambda^{n}$, $\lambda=1-\epsilon^3/2$).

\noindent iii) Finally, taking $\alpha=\kappa=0$ we find ($c={\mu \over
\beta}$):
$$(x_{n+1}x_n-1)(x_nx_{n-1}-1)=cx_n\eqno(6.5)$$
which turns out to be another form of d-P$_{\rm I}$. Its continuous limit
is obtained
through  $x=x_0(1+\epsilon^2 w)$ where $x_0^2=-1/3$, $c=-16x_0/3\lambda^{n}$,
$\lambda=1-\epsilon^5/4$ leading to $w''+3w^2+t=0$.

In order to study the degenerate cases of d-P$_{\rm V}$ it is more
convenient to go back
to the autonomous form (where $p$ and $q$ are constants):
$$(x_{n+1}x_n-1)(x_nx_{n-1}-1)={pq(x_n-a)(x_n-1/a)(x_n-b)(x_n-1/b)\over(x_n-
p)(x_n-q)}
\eqno(6.6)$$
We assume first that the numerator and denominator of the rhs of (6.6) have
one common
factor e.g. $p=a$. This gives:
$$(x_{n+1}x_n-1)(x_nx_{n-1}-1)={(1-ax_n)(x_n-b)(x_n-1/b)\over(1-x_n/q)}\eqno
(6.7)$$
In order to deautonomize (6.7) we use the singularity confinement criterion
and we find
that two solutions exist. The first corresponds to $a\propto\lambda^n$ and
$q\propto\lambda^{2n}$. In this case (6.7) is a new form of d-P$_{\rm IV}$ with
continuous limit obtained through
$x=1+\epsilon w$, $b=1+\epsilon^2 g$,
$a=-2\lambda^{n}$,
$q=4\lambda^{2n}(1-3\epsilon^2 c)$, $\lambda=1+\epsilon^2$
leading to
$w''={w'^2\over 2w}+{1\over 6}w^3-{4\over 3}w^2t+2w(t^2+c)-{2g^2\over w}$.

The second case corresponds to
$a=q\propto\lambda^n$. In this case the continuous limit is a P$_{\rm 34}$
equation
$w''={w'^2\over 2w}-{4\over 5}w^2-wt-{2g^2\over w}$
obtained through $x=1+\epsilon^2 w$, $b=1+\epsilon^3 g$,
$a=-4\lambda^{n}$, $\lambda=1-\epsilon^3/2$).

Having obtained the degenerate form (6.7) we can first perform the limit
$q\to\infty$. In
this case we have:
$$(x_{n+1}x_n-1)(x_nx_{n-1}-1)=(1-ax_n)(x_n-b)(x_n-1/b)\eqno(6.8)$$
with $a\propto\lambda^n$ as before. Equation (6.8) is still another
discrete form of P$_{\rm 34}$
(continuous limit
$w''={w'^2\over 2w}-{1\over 2}w^2-wt-{2g^2\over w}$
obtained through $x=1+\epsilon^2 w$, $b=1+\epsilon^3 g$,
$a=-3\lambda^{n}$, $\lambda=1-2\epsilon^3/3$).

The second possibility is to consider a double degeneracy where
$q=b$. We find in this case:
$$(x_{n+1}x_n-1)(x_nx_{n-1}-1)=(1-ax_n)(1-bx_n)\eqno(6.9)$$
The deautonomization of (6.9) gives $a\propto\lambda^n$,
$b\propto\lambda^n$ and the
resulting equation is a novel d-P$_{\rm II}$
(continuous limit
$w''=w^3+4wt+\sqrt{2}g$
obtained through $x=\epsilon w$,
$a=\sqrt{2}\lambda^{n}$,  $b=-(1-\epsilon^3 g)a$, $\lambda=1+\epsilon^3$).

One last limit can
be performed on this equation by taking $b=0$ while $a$ is always
proportional to
$\lambda^n$:
$$(x_{n+1}x_n-1)(x_nx_{n-1}-1)=1-ax_n\eqno(6.10)$$
This simplified equation is now a  d-P$_{\rm I}$ and its continuous limit
is given by
$x=x_0(1+\epsilon^2 w)$, $a=4x_0\lambda^{n}/3$, $\lambda=1+\epsilon^5/4$
leading to
$x''=6x^2+t$.

Before closing this section, one remark is in order. In all the cases
examined above we
found relations of d-P$_{\rm V}$ to d-P$_{\rm IV}$ and the equations
related to the latter
i.e. d-P$_{\rm 34}$ and d-P$_{\rm II}$. Thus, the question arises
naturally: ``is there
any relation of d-P$_{\rm V}$ to d-P$_{\rm III}$"? There exists of course a
Miura
transform between d-P$_{\rm III}$ and d-P$_{\rm V}$, but this is not what
we have in
mind here. (Neither do we look for a limiting process of the coalescence
type that we discussed
in section 2). From the theory of the continuous Painlev\'e equations it is
known that for a
special value of the parameters of P$_{\rm V}$ the latter reduces to a
particular P$_{\rm III}$
for some new dependent variable. Implementing this condition to our
discrete P$_{\rm V}$ we would
expect the equation:
$$(x_{n+1}x_n-1)(x_nx_{n-1}-1)={\gamma (x_n^2+1)^2\over
\alpha x_n^2+\beta x_n+\gamma}\eqno(6.11)$$
to be equivalent to some d-P$_{\rm III}$. The variable $y$ of the latter
would be
related to the variable $x$ of  d-P$_{\rm V}$ in a complicated way which, at
the
continuous limit, should reduce to $x={1\over 2}(y+{1\over y})$. However
there is no
indication as to what this transformation should be in the fully discrete
case.  Thus
the question of the relation between  d-P$_{\rm V}$ and d-P$_{\rm III}$
remains open for
the time being.
\bigskip
\noindent 7. {\scap Conclusion.}
\smallskip
\noindent The aim of this paper was to show the extreme richess of the discrete
Painlev\'e equations and of the relations that exist between them. We have
restricted
ourselves here to just the six `standard' forms of the d-P's (where the
count of six is
reached when we include d-P$_{\rm 34}$, the discrete form of d-P$_{\rm VI}$
being still
unknown). Even so, we have been able to show that many more equations than
the ones
initially obtained were `hidden' in the latter as limits or degenerate forms.

In every case examined in this paper we have only considered the non
trivial equations.
Whenever the limiting or degenerate case led to a linear equation we have
omitted it
altogether. The same was true for multiplicative equations, since the
latter can be
linearized in a straightforward way. For example, starting with
$x_{n+1}x_{n-1}=f(n)x_n$,
an equation in the d-P$_{\rm III}$ family, we can reduce it to a linear
equation by just
taking logarithms.

The analysis presented in this paper is only part of the story. As is well
known (and as
this study amply illustrates) there exist many `alternate' forms of the
d-P's. One could,
in principle, study their coalescences, limits and degeneracies as well.
Given the
sheer volume that this work represents, we prefer to leave it for some
future publication.

This study has added several new entries to the list of the discrete
Painlev\'e equations
(represented by 3-point mappings of one variable). What remains to be done
now is to
apply the arsenal we have developed in order to show that these new d-P's
have all the
special properties that characterize the Painlev\'e transcendents and which are
encountered, in a perfectly parallel way, in both continuous and discrete
equations.

\noindent
\bigskip {\scap References}.
\smallskip
\item{[1]} F.W. Nijhoff, V. Papageorgiou and H.W. Capel, Springer Lecture
Notes in Mathematics,
1510 (1992) 312.
\item{[2]}      E. Br\'ezin and V.A. Kazakov, Phys. Lett. 236B (1990) 144.
\item{[3]} B. Grammaticos and A. Ramani, {\sl Discrete Painlev\'e
equations: derivation and
properties}, NATO ASI C413 (1993) 299.
\item{[4]} G.R.W. Quispel, J.A.G. Roberts and C.J. Thompson, Physica D34
(1989) 183.
\item{[5]} A. Ramani, B. Grammaticos and J. Hietarinta, Phys. Rev. Lett. 67
(1991) 1829.
\item{[6]} P. Painlev\'e, Acta Math. 25 (1902) 1.
\item{[7]} B. Grammaticos, A. Ramani and V.G. Papageorgiou, Phys. Rev.
Lett. 67 (1991) 1825.
\item{[8]} A. Ramani, B. Grammaticos and V.G. Papageorgiou, {\sl
Singularity Confinement}, to
appear in the proceedings of the Esterel '94 conference.
\item{[9]} B. Grammaticos and B. Dorizzi, J. Math. Comp. in Sim. 37 (1994) 341.
\item{[10]} A.S. Fokas, B. Grammaticos and A. Ramani, J. Math. An. and
Appl. 180 (1993) 342.
\item{[11]} A. Ramani and B. Grammaticos, to be published.
\item{[12]} A. Ramani, B. Grammaticos and J. Satsuma, {\sl Bilinear
discrete Painlev\'e
equations}, preprint 95.
\item{[13]} A. Ramani and B. Grammaticos, Jour. Phys. A 25 (1992) L633.
\item{[14]} A. Ramani, B. Grammaticos and G. Karra, Physica A 181 (1992) 115.
\item{[15]} B. Gambier, Acta Math. 33 (1910) 1.

\end